\documentclass[12pt]{iopart}

\usepackage{epsfig}
\usepackage{iopams}
\begin{document}
\title{Antiferromagnetic spherical spin-glass model}

\author{Danilo B Liarte and Carlos S O Yokoi}

\address{Instituto de F\'{\i}sica, Universidade de S\~{a}o Paulo,
  Caixa Postal 66318, 05315-970 S\~{a}o Paulo, SP, Brazil}

\ead{danilo@if.usp.br}

\begin{abstract}
  We study the thermodynamic properties and the phase diagrams of a
  multi-spin antiferromagnetic spherical spin-glass model using the
  replica method. It is a two-sublattice version of the ferromagnetic
  spherical $p$-spin glass model.  We consider both the
  replica-symmetric and the one-step replica-symmetry-breaking
  solutions, the latter being the most general solution for this
  model.  We find paramagnetic, spin-glass, antiferromagnetic and
  mixed or glassy antiferromagnetic phases. The phase transitions are
  always of second order in the thermodynamic sense, but the spin-glass
  order parameter may undergo a discontinuous change.

\end{abstract}



\maketitle

\section{Introduction}

Mean-field theories for spin glasses have not been limited to their
original aim of explaining peculiar behaviours presented by some
magnetic alloys, but have been applied to a large number of complex
systems, ranging from biology and optimisation problems to information
processing \cite{mezard87,nishimori01}. The paradigmatic mean-field
spin-glass model is the Sherrington-Kirkpatrick (SK) model
\cite{sherrington75}. By the application of the replica method, it was
found that the low temperature spin-glass phase is described by a
solution with an infinite number of stages of replica symmetry
breaking ($\infty$RSB) according to a hierarchical scheme proposed by
Parisi \cite{parisi80}. The difficulty posed by the analysis of this
solution, analytically as well as numerically, encouraged the
investigation of simpler models that retain some of the essential
aspects of the SK model.

The generalisation from a spin-glass interaction between pairs of
spins to an interaction among sets of $p>2$ spins has attracted
considerable interest in this respect \cite{gross84}. For instance, in
the limit $p \rightarrow \infty$  this model becomes equivalent to the
random energy model (REM)\cite{derrida80}, which can be solved with or
without the help of the replica method.  Moreover, the first stage of
replica symmetry breaking (1RSB) is shown to be exact for this model.
Another instance in which the analysis becomes simpler is the
spherical version of the model \cite{crisanti92,hertz99}. This model
can be exactly solved and exhibits a spin-glass phase described by a
stable 1RSB solution for any $p$, even to the lowest temperatures.

Experimental works report evidences of a spin-glass behaviour and of
coexistence of spin-glass and antiferromagnetic orders, both in
diluted antiferromagnetic materials (e.g.
Fe${}_x$Mg${}_{1-x}$Cl${}_2$) \cite{bertrand82,wong85a,wong85b} and
in mixed antiferromagnetic compounds (e.g.
Fe${}_x$Mn${}_{1-x}$TiO${}_3$) \cite{yoshizawa87,yoshizawa94}. As a
theoretical approach to these systems, the two-sublattice SK model was
introduced to describe re-entrant transitions from the
antiferromagnetic to the spin-glass
phase\cite{korenblit85,fyodorov87}. An extended version of this model
was investigated in an attempt to reproduce the experimental results
observed in the experimental systems mentioned
previously\cite{takayama88}.  Needless to say, the solution of the
two-sublattice SK model is hard to be analysed analytically as well as
numerically. As a simpler model which retains some essential features
of the two-sublattice SK model, a two-sublattice version of the REM
was proposed recently \cite{almeida98} to explain some experimental
results in the disordered antiferromagnetic system
Fe${}_{0.5}$Zn${}_{0.5}$F${}_2$ \cite{montenegro98}. The foregoing
experimental and theoretical investigations motivated us to consider a
two-sublattice version of the spherical spin-glass model with
multi-spin interactions. The relative simplicity of the model enables
us to investigate the phase diagrams of the model for the full range
of parameters.

\section{The model}

Let us consider a set of $2N$ continuous spins distributed in two
sublattices, $A$ and $B$, each consisting of $N$ spins. The model is
defined by the Hamiltonian
\begin{equation}
\fl  \mathcal{H} = - \sum_{1\leq i_1 < \cdots < i_r \leq N \atop 1 \leq
    j_1 < \cdots < j_r \leq N}  J_{i_1 \cdots i_r
    j_1 \cdots j_r} S_{i_1} \cdots S_{i_r}\sigma_{j_1} \cdots
  \sigma_{j_r} 
  + \frac{J_0}{N} \sum_{i,j=1}^N
  S_i \sigma_j - H \sum_{i=1}^N \left(S_i+\sigma_i\right),
  \label{assgham}
\end{equation}
where $H$ is an applied magnetic field and  $J_0>0$ is the
antiferromagnetic interaction 
between different sublattices. The interactions
$ J_{i_1 \cdots i_r j_1 \cdots j_r}$ among the set of $r$ spins on
different sublattices are independent Gaussian random variables with
zero mean and variance
\begin{equation}
  \left< J_{i_1 \cdots i_r j_1 \cdots j_r}^2 \right>_J = \frac{J^2
    (r!)^2}{N^{2r-1}},
\end{equation}
where the factor $(r!)^2$ is a matter of convention while the
dependence on $N$ is needed to ensure an extensive free energy. The
spins, $S_i$ in the $A$ sublattice and $\sigma_i$ in the $B$
sublattice, are real continuous variables ranging from $-\infty$ to
$\infty$. The partition function is given by
\begin{equation}
  Z =  \int_{-\infty}^{\infty} \prod_{i=1}^N
  \rmd S_i\, \rmd \sigma_i \;
  \delta \left(N-\sum_{i=1}^N
    S_i^{\,2}\right)
  \delta \left(N-\sum_{i=1}^N \sigma_i^{\,2}\right) \rme^{-\beta \mathcal{H}},
\end{equation}
where $\beta$ is the inverse temperature and the delta functions
impose spherical constraints to ensure the existence of a well-defined
limit at low temperatures.

\section{The replica approach} 

In the replica method the self-averaged free energy per spin is
computed by means of
\begin{equation}
  -\beta f = \lim_{n \rightarrow 0}  \lim_{N \rightarrow \infty}
  \frac{1}{n N}\ln\left<Z^{n}\right>_J,
\end{equation}
where $\left<\cdots \right>_J$ denotes the average over the Gaussian
random variables $J_{i_1 \cdots i_r j_1 \cdots j_r}$ \cite{mezard87}.
A standard calculation leads to the following expression for the free
energy per spin,
\begin{eqnarray}
\fl  \beta f  = \lim_{n \rightarrow 0} \frac{1}{2n}
  \left\{ \sum_I \left[
      \sum_{\alpha,\beta}
      \left(\mathbf{q}^{-1}_I\right)^{\alpha \beta} m^{\alpha}_I m^{\beta}_I
      - 2 \beta H \sum_{\alpha} m_I^{\alpha}
      -\ln \det \mathbf{q}_I  \right] \right.
    \nonumber \\ \left. - 2\, n \ln 2\pi -\beta^2 J^2 \sum_{\alpha,\beta}
    \left(q^{\alpha \beta}_A\right)^{r}\left(q^{\alpha
        \beta}_B\right)^{r}+ 2 \beta J_0 \sum_{\alpha} m^{\alpha}_A
    m^{\alpha}_B \right\},
\end{eqnarray}
where $I=A,B$ is the sublattice index, $\alpha,\beta = 1,2, \ldots, n$
are replica indices and $( \mathbf{q}_I)^{\alpha \beta} = q_I^{\alpha
  \beta} $ with $ q_I^{\alpha \alpha} = 1$.  The saddle-point
equations for $q_I^{\alpha \beta}$ and $m_I^{\alpha}$ are given by
\begin{eqnarray}
  \sum_{\beta} \left(\mathbf{q}_{A}^{-1}\right)^{\alpha \beta}
  m_A^{\beta}- \beta H + \beta J_0 m_B^{\alpha} = 0, \\
  \left(\mathbf{q}_A^{\prime \,
      -1}\right)^{\alpha \beta} + \beta^2 J^2 r \left(q_{A}^{\alpha
      \beta}\right)^{r-1}\left(q_{B}^{\alpha
      \beta}\right)^{r} = 0, 
  \label{assgqn}
\end{eqnarray}
where
\begin{equation}
  \left(\mathbf{q}_A^{\prime}\right)^{\alpha \beta} = q_A^{\alpha \beta} -
  m_A^{\alpha} m_A^{\beta},
\end{equation}
which are coupled to similar equations obtained by the interchange $A
\leftrightarrow B$. Here, $q_A^{\alpha \beta}$ and $q_B^{\alpha
  \beta}$ denote the overlap between replicas,
\begin{eqnarray}
  q_A^{\alpha \beta}= \frac{1}{N}\sum_{i} S_{i}^{\alpha}S_{i}^{\beta},
  \qquad q_B^{\alpha \beta}= \frac{1}{N} \sum_{i}
  \sigma_{i}^{\alpha}\sigma_{i}^{\beta},
\end{eqnarray}
and $m_A^{\alpha}$ and $m_B^{\alpha}$ are the sublattice magnetisations,
\begin{eqnarray}
  m_A^{\alpha}= \frac{1}{N}\sum_{i}S_{i}^{\alpha},
  \qquad m_B^{\alpha}= \frac{1}{N}
  \sum_{i}\sigma_{i}^{\alpha}.
\end{eqnarray}

To evaluate the free energy explicitly, it is necessary to impose some
structure on $\mathbf{q}_I$ and $m_I^{\alpha}$. In this work we
consider the replica-symmetric (RS) and 1RSB {\it Ans\"atze}, the
latter being the most general solution for this model
\cite{crisanti92}. 

\subsection{The RS solution}

Usually the RS form of the overlap matrix is appropriate for the
description of systems when there is only a single equilibrium state.
We therefore expect this {\it Ansatz} to be valid in the regions of
high temperatures and high magnetic fields. Assuming for each
sublattice $I=A,B$,
\begin{equation}
  q_I^{\alpha \beta} = \left(1 - q_I\right)\delta^{\alpha \beta} + q_I,
  \qquad m_I^{\alpha} = m_I,
\end{equation}
the free energy per spin becomes
\begin{eqnarray}
  \fl   \beta f = \frac{1}{2}\sum_I \Bigg[
  \frac{m_I^2}{1-q_I} - 2 \beta H m_I -\ln \left(1-q_I\right) -
  \frac{q_I}{1-q_I}\Bigg]  - \ln 2 \pi  +  \beta J_0 m_A m_B   \nonumber \\ 
  -\frac{\beta^2 J^2}{2} \left(1-q^r_A q^r_B \right),
\end{eqnarray}
where $m_I$ and $q_I$ satisfy the saddle-point equations,
\begin{eqnarray}
  \frac{m_A}{1-q_A} - \beta H + \beta J_0 m_B = 0, \\
  \label{rsm}
  \frac{m_A^2 - q_A}{\left(1-q_A\right)^2} + \beta^2 J^2 r
  q_A^{r-1}q_B^r = 0,
  \label{rsq}
\end{eqnarray}
coupled to two similar equations obtained by the interchange $A
\leftrightarrow B$.

\subsection{The 1RSB  solution}

At low temperatures the glassy behaviour is signalled by ergodicity
breaking, with the emergence of many inequivalent pure states that are
described by the breaking of replica symmetry \cite{mezard87}.  For
the two-sublattice model the 1RSB {\it Ansatz} takes the form
\begin{equation}
  q_I^{\alpha \beta} = \left(1 - q_{1I}\right)\delta^{\alpha \beta} +
  \left(q_{1I} - q_{0I}\right)\epsilon^{\alpha \beta} +q_{0I} , \qquad
  m_I^{\alpha} = m_I,
\end{equation}
for $I=A,B$ where,
\begin{eqnarray}
  \epsilon^{\alpha \beta} = \left\{\begin{array}{ll}
      1 & \textrm{if $\alpha$ and  $\beta$
       belong to the same diagonal block}, \\
      0 &\textrm{otherwise},
    \end{array}\right.
\end{eqnarray}
which results in the following expression for the free energy per
spin
\begin{eqnarray}
  \fl  \beta f = \frac{1}{2}\sum_I \Bigg\{\frac{m_I^2}{1 -
    \overline{q}_{I}}-2 \beta H m_I -   \frac{q_{0I}}{1 -
    \overline{q}_{I}} - \ln \left(1-q_{1I}\right)
  -  \frac{1}{x}\ln \left( \frac{1 - \overline{q}_{I}}{1 -
      q_{1I}} \right) \Bigg\} 
  \nonumber \\  
  - \ln 2 \pi + \beta J_0   m_A  m_B   - \frac{\beta^2 J^2}{2}\left[1
    - \left(1-x\right)\,  q_{1A}^r\,q_{1B}^r - x\,q_{0A}^r\,q_{0B}^r \right],
\label{f1rsb}
\end{eqnarray}
where the saddle-point equations are given by,
\begin{eqnarray}
  \frac{m_A}{1-\overline{q}_A} - \beta H + \beta J_0
  m_B = 0, \\
  \beta^2 J^2 r
  q_{1A}^{r-1} q_{1B}^r +  \frac{q_{0A} - q_{1A}}{\left(1 -
      q_{1A}\right) \left(1 - \overline{q}_{A}\right)}
  -\frac{q_{0A}-m_A^2}{\left(1 - \overline{q}_{A}    \right)^2}  = 0, \\
  \beta^2 J^2 r   q_{0A}^{r-1} q_{0B}^r  - \frac{q_{0A}-m_A^2}{\left(1 - \overline{q}_{A}
    \right)^2} = 0,
\end{eqnarray}
with,
\begin{equation}
  \overline{q_I^p} = x q_{0I}^p + \left(1-x\right)q_{1I}^p,
\end{equation}
coupled to similar equations given by the interchange $A
\leftrightarrow B$. Moreover, the dimension of the diagonal blocks $x$
also contribute the additional equation
\begin{eqnarray}
  \fl   \frac{1}{2}\sum_{I} \left[\frac{\left(m_I^2 -
        q_{0I}\right)\left(q_{0I}-q_{1I}\right)}{\left(1 -
        \overline{q}_{I}\right)^2} +
    \frac{1}{x^2}\ln \left( \frac{1
        -\overline{q}_I}{1-q_{1I}} \right) +
    \frac{1}{x} \left( \frac{q_{0I} - q_{1I}}{1-\overline{q}_I} \right)
  \right]
  \nonumber \\   + \frac{\beta^2J^2}{2} \bigg(q_{0A}^r q_{0B}^r - q_{1A}^r
  q_{1B}^r \bigg) = 0.
\label{xeq}
\end{eqnarray}
Note that we have assumed the same $x$ for the diagonal blocks in 
both sublattices, because the assumption $x_A \neq x_B$ leads to the RS
solutions described by (\ref{rsm}-\ref{rsq}).

\begin{figure}[!h]
  \vspace{1.5cm}
  \begin{center}
    \epsfig{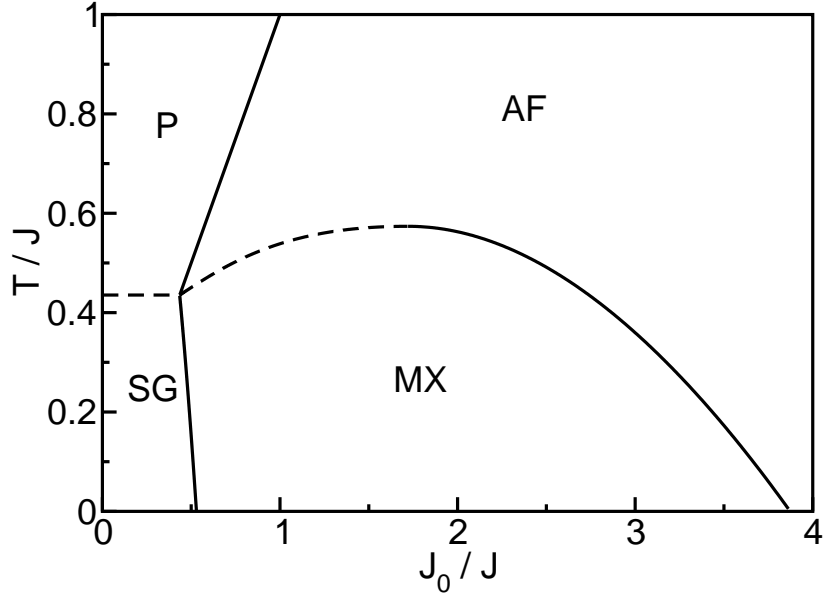}
    \vspace{0.1 cm}
    \caption{The zero-field phase diagram for $r=3$ showing the
      paramagnetic (P), antiferromagnetic (AF), spin-glass (SG) and mixed
      (MX) or glassy antiferromagnetic phases. Solid lines indicate
      continuous transitions and the dashed lines D1RSB transitions.}
    \label{figure1}
  \end{center}
\end{figure}

\section{The phase diagrams}

We present the results only for $r=3$ since the the general features
of the phase diagrams do not depend sensitively on $r$. 

\subsection{Results for $H=0$}

The zero-field phase diagram is shown in Figure \ref{figure1}.  We
found four phases which meet at a multicritical point: the
paramagnetic (P) phase, antiferromagnetic (AF) phase, spin glass (SG)
phase and mixed (MX) phase (or glassy antiferromagnetic phase).  The P
and AF phases show no replica symmetry breaking and are described by
the RS solution. The P phase is characterised by the order parameters
\begin{equation}
  m_A = m_B=0, \qquad q_A=q_B=0,
\end{equation}
and the AF phase by
\begin{equation}
  m_A = - m_B \ne 0, \qquad q_A = q_B.
\end{equation}
The SG and MX phases present replica symmetry breaking and are
described by 1RSB solution. The SG phase is characterised by the order
parameters
\begin{equation}
  m_A = m_B=0, \qquad q_{0A}=q_{0B} <  q_{1A}=q_{1B}, \qquad  0 <
  x<1 , 
\end{equation}
and the MX phase by
\begin{equation}
  m_A = - m_B \ne 0, \qquad  q_{0A}=q_{0B} <  q_{1A}=q_{1B}, \qquad  0 <
  x<1.
\end{equation}

If we make the substitutions $m_A \rightarrow m$, $m_B \rightarrow -
m$, $q_{0A}, q_{0B} \rightarrow q_0$ and $q_{1A}, q_{1B} \rightarrow
q_1$ in (\ref{f1rsb}-\ref{xeq}), these equations become identical to
those of one-sublattice $p$-spin spherical model with ferromagnetic
interactions $-J_0$ and $p=2r$ \cite{hertz99}. Thus in the absence of
an external field the results of the one-sublattice model can be
translated to the two-sublattice model simply by exchanging the
ferromagnetic and antiferromagnetic orderings.  The P-AF and the SG-MX
transitions are continuous and are characterised by the appearance of
spontaneous sublattice magnetisations $m_A = -m_B \ne 0$ in the AF and
MX phases.  The P-SG and the AF-MX transitions are characterised by
the emergence of 1RSB solution with higher free energy than the RS
solution in the SG and MX phases.  In the SG to the P transition $x
\rightarrow 1$ and $q_{1I}-q_{0I}$ vanishes discontinuously. Thus there is a discontinuous one-step
replica-symmetry-breaking (D1RSB) transition. The AF-MX transition is
also a D1RSB transition to the left of the maximum in the boundary of
MX phase, but to the right of the maximum $q_{0A}=q_{0B}$ and
$q_{1A}=q_{1B}$ merge continuously at the transition with $x < 1$.
Thus there is a continuous one-step replica-symmetry-breaking (C1RSB)
transition. We observe that although the spin-glass order parameters
change discontinuously across the D1RSB transition, thermodynamically
it is a continuous transition \cite{gross84,crisanti92}.

\begin{figure}[!h]
  \vspace{1.5cm}
  \begin{center}
    \epsfig{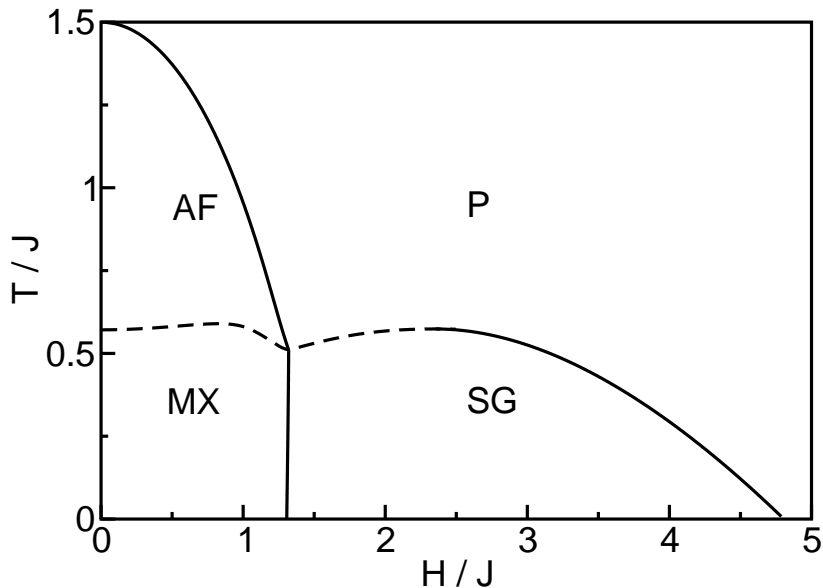}
    \vspace{0.2 cm}
    \caption{The field-temperature phase diagram of the model for $J_0/J=1.5$
      and $r=3$.  The solid lines represent continuous transitions
      and the dashed lines D1RSB transitions.}
    \label{figure2}
  \end{center}
\end{figure}

\subsection{Results for $H > 0$}

The phase diagram of the model in a uniform external field is shown in
Figure \ref{figure2} for $J_0/J=1.5$ and $r=3$. We again found four
phases that meet at a multicritical point. The P and AF phases show
no replica symmetry breaking and are described by the RS solution. The
P phase is characterised by the order parameters
\begin{equation}
  m_A = m_B \ne 0, \qquad q_A=q_B,
\end{equation}
and the AF phase by
\begin{equation}
  m_A  \ne  m_B, \qquad  q_A  \ne  q_B. 
\end{equation}
The SG and MX phases present replica symmetry breaking and are
described by 1RSB solution. The SG phase is characterised by the order
parameters
\begin{equation}
  m_A = m_B \neq 0, \qquad  q_{0A}=q_{0B}, \qquad  q_{1A}=q_{1B}, \qquad  0 <
  x<1 , 
\end{equation}
and the MX phase by
\begin{equation}
  m_A \ne  m_B , \qquad  q_{0A} \ne q_{0B}, \qquad  q_{1A} \ne q_{1B}, \qquad  0 <
  x<1.
\end{equation}

\begin{figure}[!h]
  \vspace{1.5cm}
  \begin{center}
    \epsfig{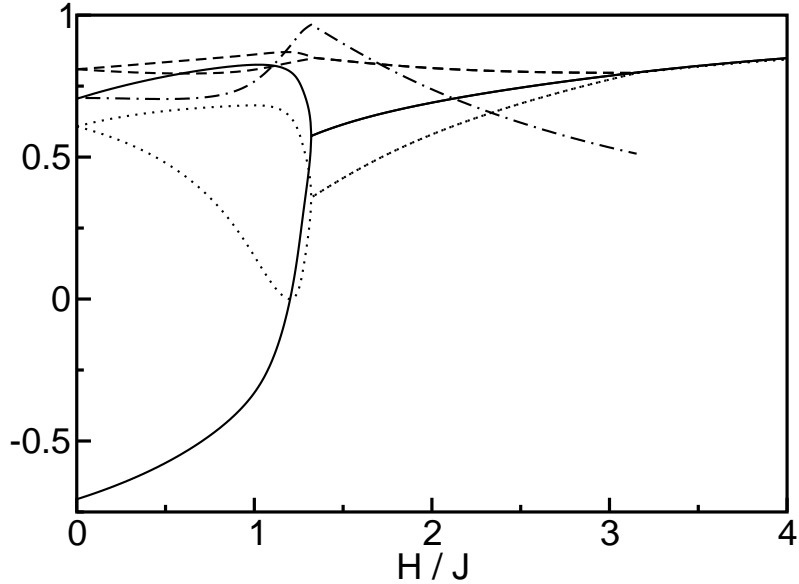}
    \vspace{0.2 cm}
    \caption{The order parameters as functions of the magnetic field
      for $r=3$, $J_0/J=1.5$ and $T/J=0.5$. The full lines represent
      $m_A$ and $m_B$, the dotted lines $q_{0A}$ and $q_{0B}$, the
      dashed lines $q_{1A}$ and $q_{1B}$, and the dash-dotted line
      represents $x$.}
  \label{figure3}
  \end{center}
\end{figure}
\begin{figure}[!h]
  \vspace{1.5cm}
  \begin{center}
    \epsfig{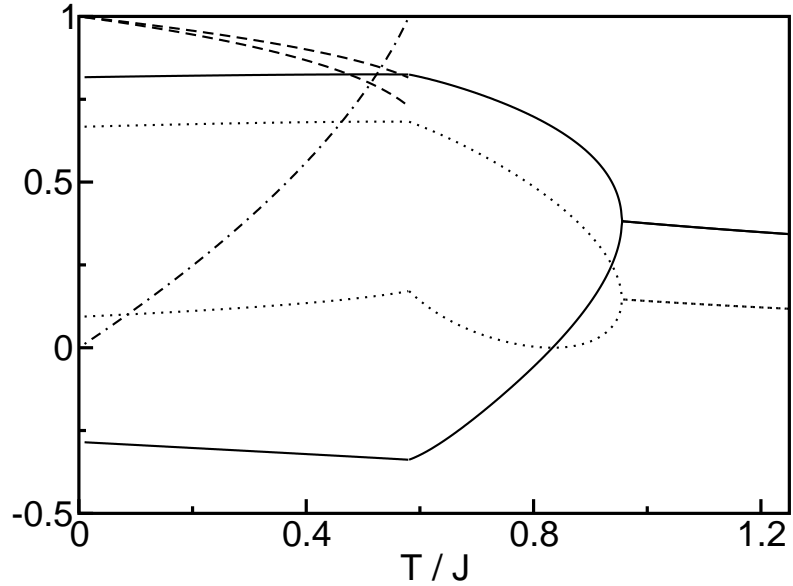}
    \vspace{0.2 cm}
    \caption{The order parameters as functions of the temperature for
      $r=3$, $J_0/J=1.5$ and $H/J=1$.  The full lines represent
      $m_A$ and $m_B$, the dotted lines $q_{0A}$ and $q_{0B}$, the
      dashed lines $q_{1A}$ and $q_{1B}$, and the dash-dotted line
      represents $x$.}
  \label{figure4}
  \end{center}
\end{figure}

The order parameters as functions of the external field is plotted in
Figure \ref{figure3} for $r=3$, $J_0/J=1.5$ and $T/J=0.5$, when the
system undergoes transitions from MX to SG and from SG to P phases for
increasing fields. The MX-SG transition takes place at $H/J = 1.32$.
At this transition the sublattice order parameters $m_A$ and $m_B$,
$q_{0A}$ and $q_{0B}$, and $q_{1A}$ and $q_{1B}$ become identical with
$x<1$, implying a continuous transition between RSB phases.  At the
SG-P transition at $H/J = 3.15$ the order parameters $q_{0A}=q_{0B}$
and $q_{1A}=q_{1B}$ merge continuously with $x < 1$, indicating a
C1RSB transition.

Figure \ref{figure4} shows the order parameters as functions of the
temperature for $r=3$, $J_0/J=1.5$ and $H/J=1$, when the system
undergoes transitions from MX to AF and from AF to P phases for
increasing temperatures. The MX-AF transition takes place at
$T/J=0.58$. At this transition the sublattice magnetisations $m_A$ and
$m_B$ meet continuously. However $q_{0B}$ and $q_{1B}$ do not merge
continuously with $q_{1A}$ and $q_{1B}$. Also $x \rightarrow 1$ at the
transition, which indicates a D1RSB transition.  The AF-P transition
at $T/J = 0.95$ is a conventional continuous transition between RS
phases.

\begin{figure}[!h]
  \vspace{1.5cm}
  \begin{center}
  \epsfig{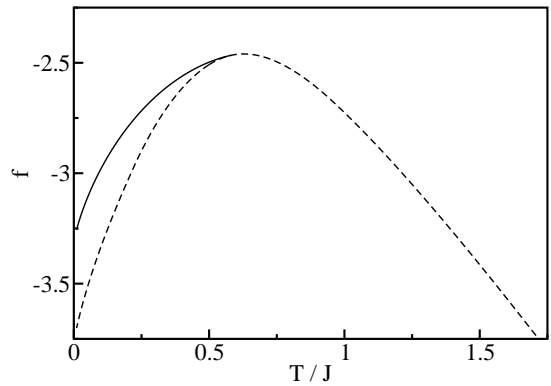}
  \caption{Free energy per spin as a function of temperature for
    $H/J=1.0$, $J_0/J=1.5$ and $r=3$.  The solid and dashed lines
    represent the 1RSB and RS solutions respectively.}
  \label{figure5}
 \end{center}
\end{figure}

The free energies of the RS and 1RSB solutions for $r=3$, $J_0/J=1.5$
and $H/J=1$ are shown in Figure \ref{figure5} across the MX-AF
transition.  We observe that in the MX phase the free energy of the
1RSB solution is higher than the RS solution.  A similar behaviour was
also found in the one-sublattice model \cite{crisanti92}. In the
replica approach this indicates that we must choose the 1RSB rather
than the RS solution \cite{mezard87}.

\begin{figure}[!h]
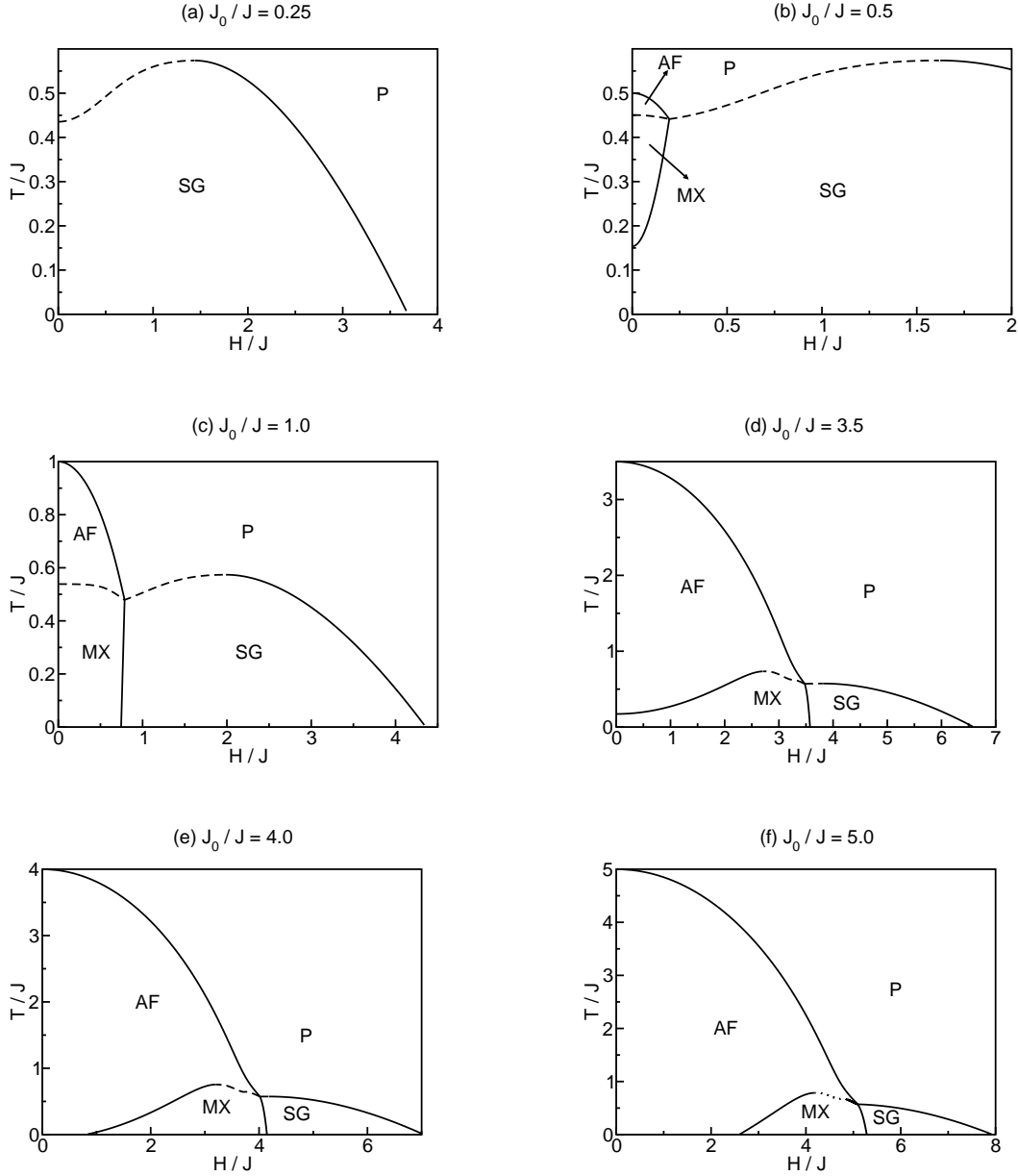

  \vspace{1.5cm}
  \begin{minipage}[t]{0.50\linewidth}
    \epsfig{file=./figure6a.eps,scale=0.25}
  \end{minipage}
  \begin{minipage}[t]{0.50\linewidth}
    \epsfig{file=./figure6b.eps,scale=0.25}
  \end{minipage}
  \begin{minipage}[t]{0.50\linewidth}
    \vspace{0.35 cm}
    \epsfig{file=./figure6c.eps,scale=0.25}
  \end{minipage}
  \begin{minipage}[t]{0.50\linewidth}
    \vspace{0.35 cm}
    \epsfig{file=./figure6d.eps,scale=0.25}
  \end{minipage}
  \begin{minipage}[t]{0.50\linewidth}
    \vspace{0.85 cm}
    \epsfig{file=./figure6e.eps,scale=0.25}
  \end{minipage}
  \begin{minipage}[t]{0.50\linewidth}
    \vspace{0.85 cm}
    \epsfig{file=./figure6f.eps,scale=0.25}
  \end{minipage}
  \vspace{0.1 cm}
  \caption{The field-temperature phase diagrams for various values of
    $J_0$. The full line represent continuous transitions while the
    dashed and dotted lines represent D1RSB transitions.}
  \label{assghtj0}
\end{figure}

Figures \ref{assghtj0}(a)-(f) show the evolution of the
field-temperature phase diagrams for increasing values of the
antiferromagnetic coupling $J_0$. In Figure \ref{assghtj0}(a) for
$J_0/J=0.25$ only the SG and the P phases are present.  The P-SG
boundary has a maximum. It is a D1RSB transition to the left of
maximum and C1RSB transition to the right. This feature of P-SG
transition is common to all the subsequent phase diagrams.  In Figure
\ref{assghtj0}(b) for $J_0/J=0.5$ the AF and MX phases are present for
small magnetic fields.  The P-AF and MX-SG transitions are always
continuous. The MX-AF transition is of D1RSB type.  In Figure
\ref{assghtj0}(c) for $J_0/J=1.0$ the MX phase extends to zero
temperature. In Figure \ref{assghtj0}(d) the AF-MX boundary exhibits a
C1RSB transition in the low field side. This feature of the AF-MX
transition which starts at $J_0/J = 1.72$ is common to all subsequent
phase diagrams. In Figure \ref{assghtj0}(e) for $J_0/J=3.5$ the MX
phase is no longer present at zero field.  Finally, in Figure
\ref{assghtj0}(f) the D1RSB transition line has disappeared in the
P-SG boundary, and only the C1RSB transition line remains.  Also the
the D1RSB transition in the MX-AF boundary becomes again a C1RSB
transition before reaching the multicritical point.

\section{Conclusions}

We have studied the thermodynamic properties and phase diagrams of a
$r$-spin antiferromagnetic spherical spin-glass model using the
replica method. For this class of models the first step in the replica
symmetry breaking is sufficient \cite{crisanti92}. The model is a
two-sublattice version of the $p$-spin ferromagnetic spherical
spin-glass model \cite{hertz99}.  The two models become essentially
identical in their properties in the absence of an external field if
the roles of the ferromagnetic and the antiferromagnetic orderings are
exchanged.  The model can also be considered as a spherical version of
the antiferromagnetic SK model
\cite{korenblit85,fyodorov87,takayama88} and the antiferromagnetic REM
model \cite{almeida98}.

We have presented a detailed numerical study for the representative
case $r=3$.  The phase diagrams comprise the paramagnetic (P) phase,
the antiferromagnetic (AF) phase, the spin-glass (SG) phase and mixed
(MX) or glassy antiferromagnetic phases. All the transitions between
these phases are continuous in the thermodynamic sense. However the
spin-glass order parameters may change continuously (C1RSB) or
discontinuously (D1RSB) in the SG-P and MX-AF transitions. 

Previous studies of the same problem in antiferromagnetic SK model
\cite{korenblit85,fyodorov87,takayama88} and the antiferromagnetic REM
model \cite{almeida98} have yielded qualitatively similar phase
diagrams. However the P-SG and AF-MX transitions are of continuous
$\infty$RSB type in the antiferromagnetic SK model
\cite{korenblit85,fyodorov87,takayama88} and of continuous C1RSB
type in the REM model \cite{almeida98}.

\section*{Acknowledgement}

D. B. Liarte acknowledges the financial support from Conselho Nacional
de Desenvolvimento Cient\'ifico e Tecnol\'ogico (CNPq).

%
%
\end{document}